\documentclass{raa}

\usepackage{graphicx,times}             %for PS/EPS graphics inclusion, new
\input{epsf.sty}                        %for PS/EPS graphics inclusion, old
\input{psfig.sty}                       %for PS/EPS graphics inclusion, old

\begin{document}

   \title{Possible Origin of the Damocloids:the Scattered Disk or a New Region?
%\,$^*$
%\footnotetext{$*$ Supported by the National Natural Science Foundation of China.}
}
%   \subtitle{I. Place Your Subtitle Here}

   \volnopage{Vol.0 (200x) No.0, 000--000}      %%preserved for Editor. DOn't remove!
   \setcounter{page}{1}          %%starting page, preserved for Editor. DOn't remove!

   \author{S. Wang
      \inst{1}
   \and H. B. Zhao
      \inst{1}
   \and J. H. Ji
      \inst{1}
   \and S. Jin
      \inst{1,2}
   \and Y. Xia
      \inst{1}
   \and H. Lu
      \inst{1}
   \and M. Wang
      \inst{1}
   \and J. S. Yao
      \inst{1}
   }
%% Here is an example of three authors come from different institutes.
%% For single author or all the authors from an institute, use "\inst{}" only

   \institute{Purple Mountain Observatory, Chinese Academy of Sciences,
Nanjing 210008, China; {\it jijh@pmo.ac.cn}\\
%% Please give the E-mail address of the author, to whom future correspondence and
%% offprint requests will be sent.
        \and
             Graduate School of Chinese Academy of Sciences,
Beijing 100049, China\\
           }

   \date{Received~~2011 August 30; accepted~~2011~~month day}

\abstract{The Damocloids are a group of unusual asteroids, recently
enrolling a new member of 2010 EJ104. The dynamical  evolution for
the Damocloids may uncover a connection passage from the Main Belt,
the Kuiper Belt and the scattered disk beyond. According to our
simulations, two regions may be considered as possible origin of the
Damocloids: the scattered disk, or a part of Oort cloud which will
be perturbed to a transient region locating between 700 AU and 1000
AU. Based on the potential origin, the Damocloids can be classified
into two types, with relation to their semi-major axes, and about
65.5\% Damocloids is classified into type I which mainly originate
from Oort cloud. Whether the Damocloids is inactive nuclei of Halley
Family Comets may rely on their origin.  }

   \authorrunning{Wang et al.}            %author_head in even pages
   \titlerunning{Possible Origin of the Damocloids}  % title_head in odd pages

   \maketitle
%% The author head (on even pages) and the title head (on odd pages) will be
%% automatically extracted from \author{} and \title{}. Whenever the title is too long,
%% you will be asked to supply a shorter one by inserting either \authorrunning{} or
%% \titlerunning{} before \maketitle. Anyway, you can specify your own heads.
%%
%%
%% Note: In the following text body of your manuscript, please note several differences from
%%       other major journals:
%% (1) \subsection{Please Capitalize the First Letter of Each Notional Word in Subsection Title}
%% (2) Please Capitalize the First Letter of Each Notional Word in all tables' captions

%
%________________________________________________ sections below
%
\section{Introduction}           %% first-level sections will be auto-capitalized
\label{sect:intro}

In our solar system, most asteroids are classified into following
types: I) Main Belt asteroids, near-earth objects and Jupiter
trojans, mostly reside in the inner solar system; II) Centaurs
mainly travel in the outer solar system between the regime of
Jupiter and Neptune; III) Trans-Neptune objects are at or beyond the
orbit of Neptune, e.g., Neptune trojans, Kuiper Belt objects and
scattered disk objects. Different from asteroids, there are comets
which have visible signs of outgassing in the solar system,
containing Jupiter-family comets (JFCs) and Halley-family comets
(HFCs). JFCs are in short periods less than 20 years, crossing
Jupiter's orbit and dynamically dominated by major planets (\cite
{Duncan08}). HFCs are comets with orbital periods between 20$\sim$200
years, with perihelion distances less than 1.5 AU (\cite {BE96}).

To date, there are many other small bodies which cannot be
catalogued into the aforementioned types, e.g., 1998 WU24 ($e=0.9$
,$q=1.4$ AU) (\cite{Davies01}) and 1999 LD31 ($e=0.90$, $q=2.38$ AU)
(\cite{Harris01}), both moving on an eccentric orbit. Both of them
share similar orbits with HFCs, without visible signs of outgassing.
According to their common properties, these bodies now belong to a
new population - the Damocloids (\cite{Jewitt05}).

The Damocloids are objects which have a Tisserand Parameter with
respect to Jupiter not larger than 2 (\cite{Jewitt05}). The Tisserand
Parameter is expressed as,
\begin{equation}
T_J=\frac{a_J}{a}+2[(1-e^2)\frac{a}{a_J}]^{1/2}\cos i,
\label {TP}
\end{equation}
where $a_J,~a,~e,~i$ refer to the semi-major axis of Jupiter, the
semi-major axis, eccentricity and inclination of the small body,
respectively. According to the definition, there are 77 Damocloid
candidates, consisting of 41 Damocloids with decent orbits up to
2011 February. The Tisserand Parameter is always used to distinguish
the JFCs and HFCs and other asteroids in the solar system. For JFCs,
we have $2<T_J<3$; while for HFCs we have $T_J<2$, which is also the
criteria for the Damocloids. But for other asteroids, one may obtain
$T_J>3$.

On March 10, 2010, we discovered a new asteroid moving along a very
eccentric orbit, designated as 2010 EJ104. We observed this rapidly
moving object using 1.04/1.20 m Schmidt Telescope (Near Earth Object
Survey Telescope) at Xuyi station of Purple Mountain Observatory
(PMO) (\cite{Zhao07, Zhao09, Zhao10}). The orbital elements are then
determined by utilizing follow-up measurements from several
observatories -- the semi-major axis $a=21.58$ AU, eccentricity
$e=0.90$, and inclination $i=41.55^\circ$, perihelion distance
$q=2.13$ AU. The orbital elements and their 1 $\sigma$ uncertainties
are shown in Table 1. The orbital feature shows that this object is
similar to 1998 WU24 and 1999 LD31 (\cite{Davies01,Harris01}). On the
basis of the orbital data, the Tisserand Parameter for 2010 EJ104 is
about 1.568. Additionally, the object is similar to other members of
the Damocloids, which are also no cometary feature. Hence, the
Damocloids now enroll a new member of 2010 EJ104, with a total
number up to  42.

Then a natural question may be raised -- where does the Damocloids
come from and is there a source or region that may replenish these
comparable objects? As they have similar orbital properties with
HFCs, if they arise from same source, or in other words, are the
Damocloids  the inactive nuclei of HFCs? Nowadays, several scenarios
have been proposed to shed light on possible dynamical origin for
such objects.

Firstly, minor objects in the inner solar system could be scattered
out due to severe perturbations by giant planets (e.g., Jupiter and
Neptune) during the secular evolution in the planetesimal disks
(\cite{Ray04, Ji05, Ji11}). For example, the trans-Neptunian objects
could undergo the process of being scattered inward by Neptune
(\cite{Levison09, MM}). The trans-Neptunian objects is currently
postulated to originate from the Kuiper Belt and scattered disk
(\cite{Glad05, ML}). Recently, the dust disk in the Kuiper Belt may
also serve as an alternate origin for these objects. Additionally,
the nearby region of Jupiter may be considered as the birthplace of
such objects, where $\sim 8\%$ objects residing the orbit between
Jupiter and 3.3 AU are ejected to an eccentric orbit in the
simulations (\cite{WL97}).

Secondly, the objects in the outer solar system that could be
perturbed by passing massive stars or tidal effect of the Galactic
disk. In the early 1950's, a spherical cloud, consist of numerous
comets ranging from 2,000 AU to 50,000 AU, was firstly postulated to
mark up the distant barrier of the solar system (\cite{oo,Wei,Don}),
which was known as so-called Oort Cloud. Generally speaking, the Sun
and major planets play insignificant part in bringing about such
bodies due to their faraway distance. However, the orbital stirring
arising from the passing stars or massive planets are quite
significant (\cite {Mor08, LM04}) and the tidal effects exerted by the
disk (\cite {Fou06}) and bulge of the Milky Way may still play a very
important part in producing these small bodies (\cite{Byl90}).
Suffering from complicated dynamical process, the objects in the
Oort cloud will be greatly excited and further driven to be thrown
into the inner solar system over longer timescale of evolution.

Recently, an alternative mechanism was proposed to explain the
origin of small objects. In this scenario, a hypothetical Sun's
companion (\cite{Matese}), with a mass of several Jovian masses
wandering about the innermost region of the outer Oort cloud, may
induce the detached Kuiper Belt objects to migrate inward and then
cross over the disk, and eventually close approach the Sun.
Consequently, the companion may probably trigger transportation of
the objects like the Damocloids.

According to close similarity, the Damocloids may be inactive nuclei
of HFCs (\cite {Jewitt05,toth06}) with same origin, while there are
two main sources of HFCs: the scattered disk (\cite{Levison06}) and
the inner Oort cloud (\cite{Levison01}) locating from 2000 AU to 20000
AU. Under the gravitational effect of the Sun, eight planets,
passing stars, and Galactic tides, with proper distribution of
inclination about 50$^{\circ}$, the inner Oort cloud objects are
stirred up to give birth to HFCs. According to the scattered disk
model, it may predict that the number of the HFCs will be roughly 10
times of the currently observed results, and the possible outer
boundary of the scattered disk is about 200 AU.

In this Paper, we extensively study a general origin of the
Damocloids through numerical simulations. And we find a transient
disk locating from 700 AU to 1000 AU. Before the object become a
Damocloids, it may pass through this region. In $\S 2$, we briefly
introduce the method, to numerically investigate the dynamical
origin of the Damocloids. Next, we analyze simulation results and
discuss the origin mechanisms in $\S 3$ , and we summarize the
outcomes in $\S 4$.

\begin{table*}[htp]
\begin{center}
\caption{Heliocentric ecliptical orbital elements of 2010 EJ104 for
TDB epoch 2455400.5 (Reference frame is ICRF/J2000)\label{tbl-1}}
\begin{tabular}{crr}
  \hline\noalign{\smallskip}
Elements          & Values                 & Uncertainty           \\
  \hline\noalign{\smallskip}
$a$ (AU)          & 21.586266636538866     & 5.688461074865734E-2  \\
$e$               & 0.9011439847926949     & 2.5546023114618503E-4 \\
$i$ (deg)         & 41.551986960824294     & 3.1068112777202545E-3 \\
peri (deg)        & 177.04115294493582     & 7.3861295389317065E-3 \\
node (deg)        & 353.0291168885478      & 3.357376430876397E-4  \\
$t_P$ (JD)        & 2455271.4918569964     & 1.7166562491213885E-2 \\
$M$ (deg)         & 1.2678120987288926     & 4.901376431292673E-3  \\
$q$ (AU)          & 2.1339323028906296     & 1.1173156970862058E-4 \\
  \hline\noalign{\smallskip}
\end{tabular}
\end{center}
\end{table*}

%% Authors can give a citation as 'Michel et al. 1992'.
%% You may also use \cite, \citep and \citet for citation, and use Table~1 or Figure~1
%% and so forth. Using \ref and \label for cross-references of Tables/Figures
%% is a good way in adjusting/adding/removing text, tables or figures.

\begin{figure}
% \figurenum{5}
    %%\epsscale{1.30}
\includegraphics[scale=0.7]{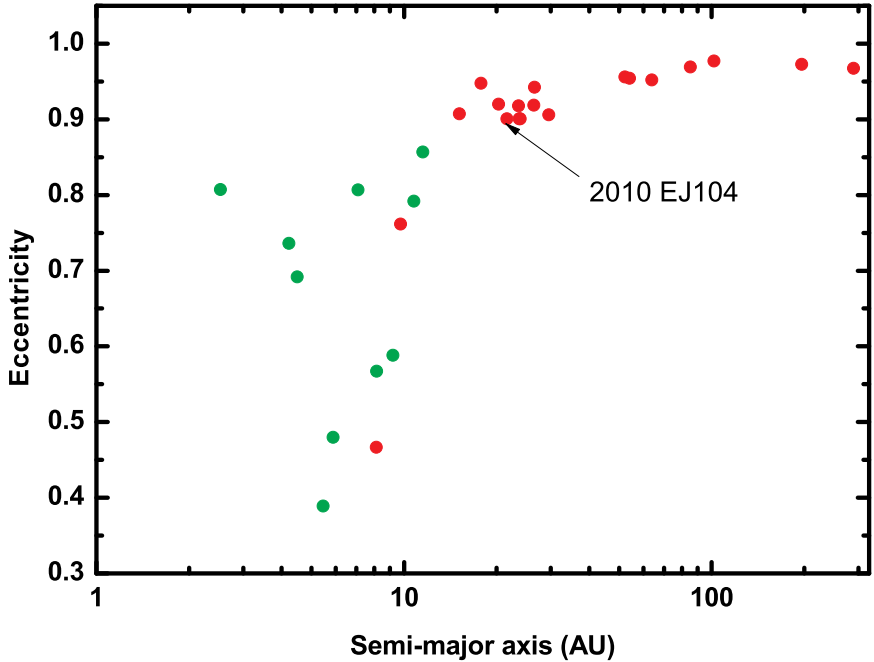}
  \caption{The semi-major axis versus eccentricity distribution of 29 Damocloids,
  where 2010 EJ104 is marked out in the figure. According to their possible
  origin, 29 Damocloids can be classified into four types: Type I and II are labeled in red, pink, yellow and green, respectively.}
 \label{fig1}
\end{figure}

\begin{figure}
% \figurenum{3}
   \plotone{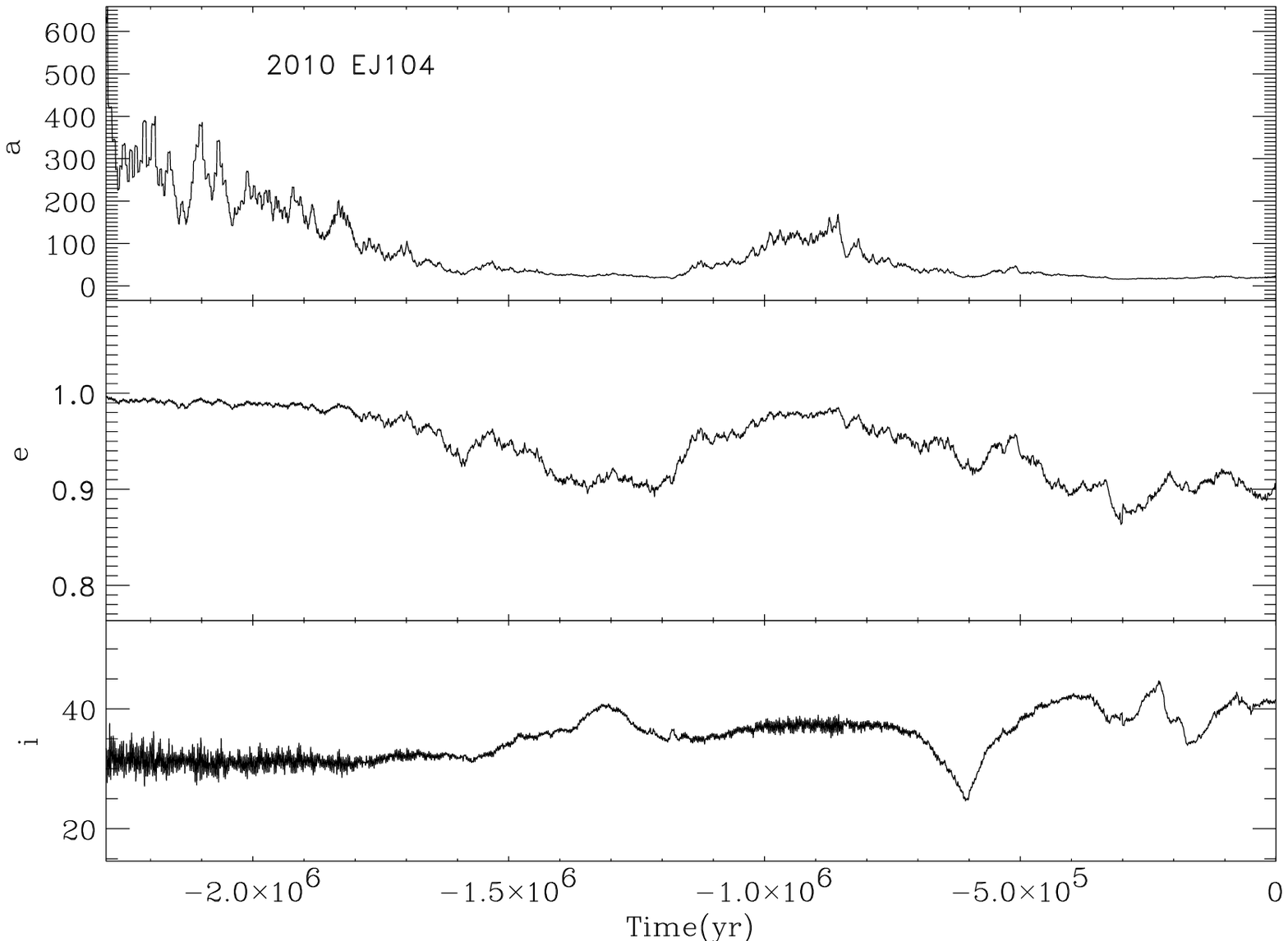}
  \caption{The orbit of 2010 EJ104 is integrated backwards about 2 Myr.
 Three panels shows the dynamical evolution of its semi-major axis $a$,
 eccentricity $e$ and inclination $i$, respectively. At the end of the simulation,
 the object transfers to hyperbolic orbit at $\sim$ 600 AU.}
 \label{fig2}
\end{figure}
\section{Methods}

To better understand the possible dynamical origin of the
Damocloids, we investigate the past motion of them by tracing back
the orbits under the perturbation of the main planets in the solar
system. Backwards integration will illustrate the orbit before the
Damocloids moving to the position that we observed. Herein we carry
out the backwards numerical simulations in a heliocentric system
using a hybrid symplectic algorithm in the MERCURY package
(\cite{Cham99}).

We carry out 29 runs with the 29 target objects shown in Figure
\ref{fig1} from the Damocloids family including 2010 EJ104. In order
to derive the original orbit, we creates a swarm of test particles
around each target object with their nominal orbital element values
we observed. The initial semi-major axes, eccentricities and
inclinations of the test particles are induced randomly in the range
of the element uncertainties. In addition, the other initial orbital
elements of each test particle are randomly generated - the
arguments of periastron, longitudes of the ascending node, and mean
anomalies range from $0^{\circ}$ to $360^{\circ}$. For each run, we
integrate backwards over the timescale of $10^8$ years.
Additionally, the hybrid integrator parameters are adopted as a
stepsize of $6$ days, a Bulirsch-Stoer tolerance of $10^{-12}$. In
all runs, the gravitational interaction of the Sun and eight major
planets in the solar system are fully taken into account during the
integration. We stop the calculations when the test particles reach
the distance from the central star larger than 1000 AU which is out
of the perturbation of the inner solar system (mainly the effect of
the eight major planets and the sun). On the other hand, when the
test particle run into the distance closer than the radius of the
sun, we assume that it collide with the sun.

At the end of the simulation, we record the information when one of
the conditions is satisfied: I) the orbit of the test particle is
changed into hyperbolic; II) the distance from the test particle to
the sun is larger than 1000 AU; III) the distance from the test
particle to the sun is closer than the radius of the sun. Above
three conditions are probably caused by the reasons following, I)
perturbation from the main planets in the solar system; II) escaping
from the inner solar system (here we define $d<1000$ AU as the inner
solar system, $d$ means the distance from the central star); III)
glancing from the sun. Thus, from analysis the record, we will get
the possible origin of the Damocloids.
\begin{figure}
% \figurenum{4}
\centering
   \includegraphics[scale=0.8]{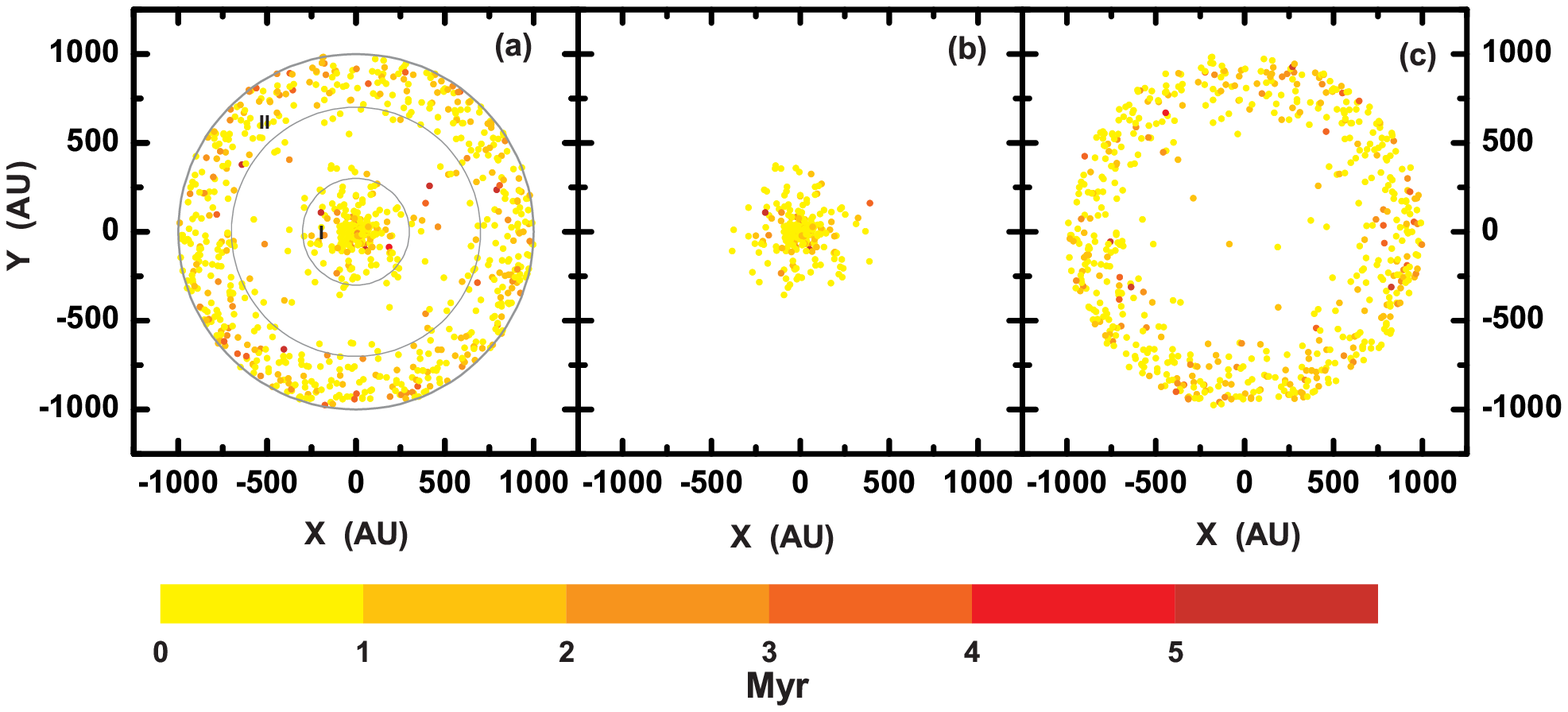}
   \vspace{-5cm}
  \caption{Results of  backward simulations. Based on observational uncertainties,
 all test bodies are initially set to be 21.586 $\pm$ 0.057 AU. In the figure,
 all of bodies had been tracked for their birthplace. The bodies from the disk within
 300 AU is marked up as Region I (the outer belt), while those from
 700 $\sim$ 1000 AU are labeled Region II (inside the innermost circle). The average lifetime of all  bodies is
 about $9.93 \times 10^5$ years, where deeper color index indicates longer lifetime. Panel (b) shows the test particles whose orbits changed into hyperbolic or collide with
 the sun.
  Panel (c) shows the test particles which escaped from the inner solar system.}
 \label{fig3}
\end{figure}

\begin{figure}
% \figurenum{4}
\centering
   \includegraphics[scale=0.6]{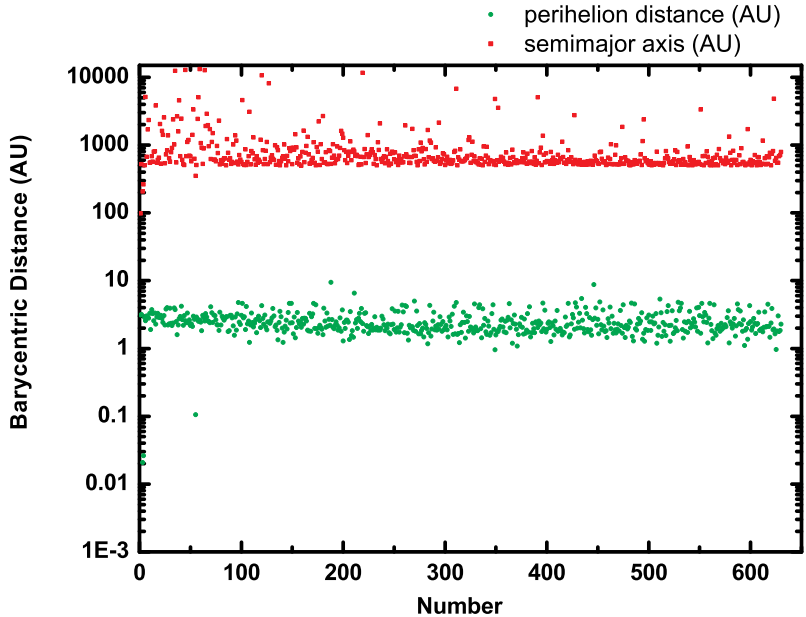}
      \vspace{-1cm}
  \caption{Statistic results of the objects which run back to Region II in the simulation of 2010 EJ104. The red filled squares mean the
  semimajor axes and the green points represent the perihelion distance. The grey dot lines display the location of the main
planets from Mars to Neptune.}
 \label{fig4}
\end{figure}

\begin{figure}
% \figurenum{4}
\centering
   \includegraphics[scale=0.6]{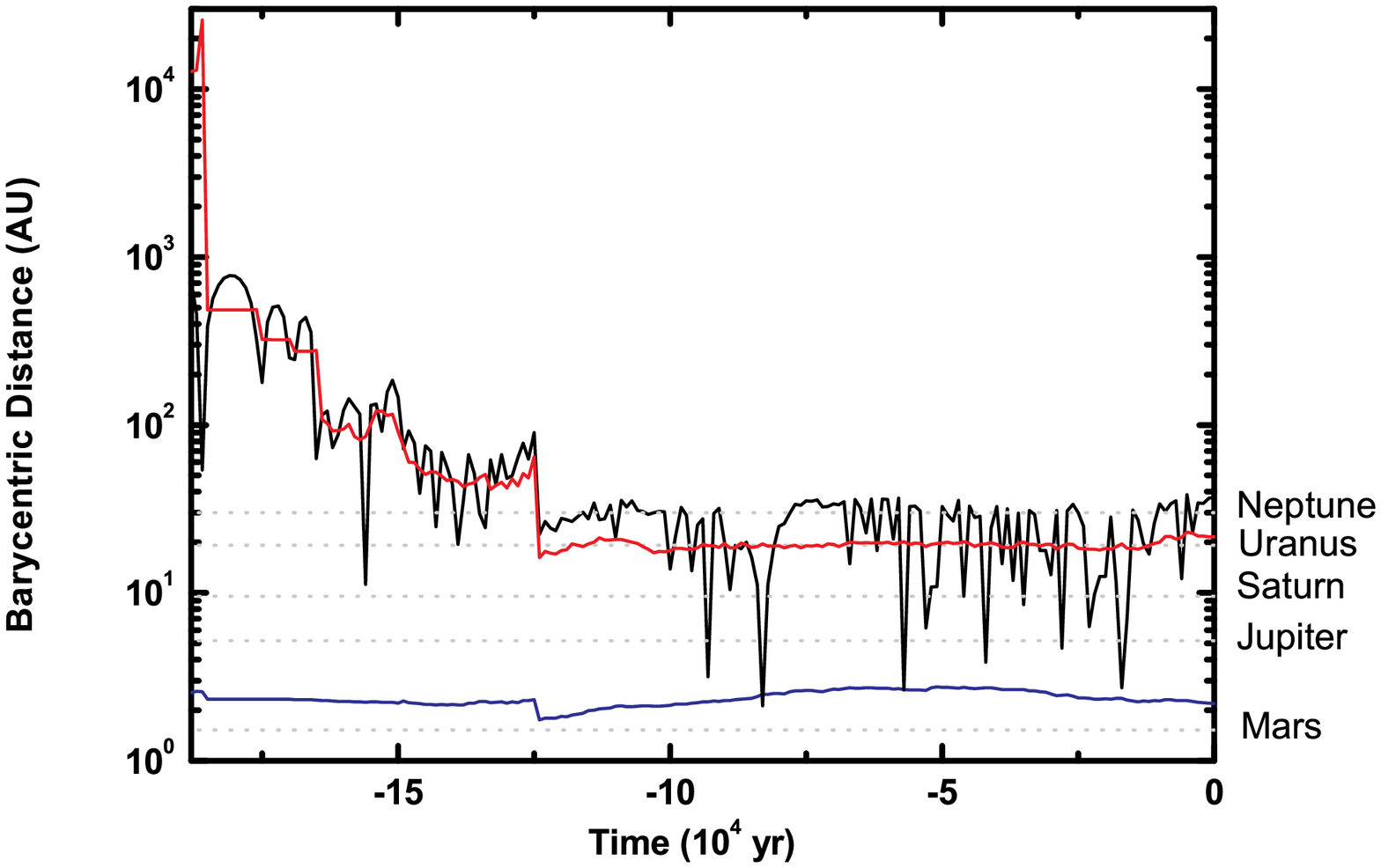}
         \vspace{-0.5cm}
  \caption{Typical result of the object which run back to Region II. The red line means
  the evolution of semimajor axis, blue line represents the perihelion distance and the black
  line shows the evolution of the barycentric distance.}
 \label{fig5}
\end{figure}

\section{The Scenario of Origin}

Now we come to the simulation results. We analysis the results
taking 2010 EJ104 as an example. Figure \ref{fig2} shows a typical
run of 2010 EJ104. In this run, the initial parameters are adopted
from Table 1. Eventually, the orbit of this object changes to
hyperbolic with a dynamical timescale of $\sim 2$ Myr at  $\sim 600$
AU, where it may originate from.

Figure \ref{fig3} shows the orbit information of each test particle
for 2010 EJ104 at the end of the simulation. From panel (a) of this
figure, it can be noticed that most of test particles in our
simulations run back to mainly two regions -- the disk at a distance
within 300 AU away from the Sun (labeled Region I) or a disk between
700 AU and 1000 AU (labeled Region II). Analysis the results, we
find that, Region I is mainly composed by two kinds of test
particles. One's orbits has been changed into hyperbolic at the end
of the simulation as shown in panel (b) of figure 3 and another
collides with the sun in the backward simulation. While the test
particles in Region II are almost the ones which escaped from the
inner solar system with $d>1000$ AU at last, the distribution of
such test particles are shown in panel (c). Thus, there are two
kinds of test particles in Region I. Considering the hyperbolic
orbit at the end of the simulation, One part may be involved in the
scattering scenario due to stirring by major planets when it lies in
the scattered disk (\cite{DL97}) or the Main Belt, especially the
scattered disk will encounter Neptune during the dynamical evolution
in solar system (\cite{Glad05}). Another part is those that may be
originated from the glancing of the sun. 37\% test particles in
Region I.

The objects in Region II will escape to the outer solar system
further than 1000 AU if we continue our calculation. Thus the
objects in Region II may be attributed to Oort cloud by three
mechanisms as aforementioned: the first one is the influence of a
passing star or massive planets (\cite {Mor08}), the second
mechanism is the induced tidal effect by the galactic disk (\cite
{Fou06}), and the third scenario may result from the perturbation of
the solar companion (\cite{Matese}). Although only the Galactic
tides cannot bring the objects from the inner Oort cloud to the
locations below $10^3$ AU, about 70\% of them with small perihelion
can be brought to inner region of solar system due to the
perturbation of the main giant planets (\cite{Levison01}). From the
results of Levison et al. 2001, if the object in the inner Oort
cloud with the perihelion distance around the locations of the main
giant planets in the solar system, gravitational encounters with the
giant planets cause random walk in semimajor axis. If semimajor axis
decreases to about 30 AU, outer giant planet will hand off it to the
inner giant planet and finally scatter it into the inner solar
system. If semimajor axis increases to distance larger than 15000
AU, the effect of Galactic tide becomes significant and will take
the object to a orbit with few hundreds AU semimajor axis and small
perihelion distance. Another evolution process, a passing star can
lower the perihelion distance of the object in the inner Oort cloud
to the inside of Jupiter's orbit. Then the interaction with Jupiter
will make it to the region locating from 700 AU to 1000 AU. Figure
13 of their paper shows four kinds of evolution processes from the
inner Oort cloud to the inner solar system. Figure 4 in our paper
shows the semimajor axes and perihelion distances results of all the
objects that run back to Region II at the end of the simulations. In
order to explain our results clearly, we change our results to
barycentric system in figure 4 and 5. The perihelion distances shown
by the green points in figure 4 are mostly range from 1 AU to 10 AU
where can be perturbed by the main planets. Figure 5 shows a typical
evolution process. The red line, blue line, and the black line
represent the evolution of the semimajor axis, the perihelion
distance, and the barycentic distance respectively. The grey dot
lines display the location of the main planets from Mars to Neptune.
The object runs back to the barycentric distance about $10^3$ AU.
The forthcoming evolution process can be described as follows.
Firstly, object from the inner Oort cloud is perturbed to the Region
II. Then close encounter with Neptune will hand off the object to
Saturn and in turn to Jupiter. Finally interacting with Jupiter
makes it a Damocloids. According to this formation scenario, before
object which come from Oort cloud become Damocloids, it must transit
the disk. From the statistic result, 63\% test particles run back to
Region II.

%As the Kuiper Belt ranges from Neptune up to 500 AU away from the
%Sun, the quasi-static disk between 500 $\sim$ 800 AU we assumed may
%be at the outer edge of the Kuiper Belt. Hence, the total mass of
%objects in the quasi-static disk is related to the Oort cloud. The
%mass of the Oort cloud is estimated to be five Earth-mass
%(\cite{Mor08, Wei83}) and might be amount up to 380 Earth-mass in
%the earlier stage. Recently, a new insight indicates that over 90\%
%comets in the Oort cloud may be captured from other stars when they
%are in the birth cluster (\cite{levison10}). According to this
%model, the number of the comets in Oort cloud may reach $4 \times
%10^{11}$ with radius larger than 1 km. According to the mass of a
%typical comet $4 \times 10^{16}$ g, the Oort cloud may contains 2.67
%$M_{\oplus}$ locating from 2000 AU to 50000 AU. Meanwhile, the
%number of the scattered disk is suggested to be $\sim 6 \times 10^8$
%with a mass of about 0.004 $M_{\oplus}$ at 30 AU to 100 AU relying
%on observational constraints on the JFCs and dynamical simulations
%(\cite {levison97}). Considering the mass and the distance
%relationship in Oort cloud and scattered disk, we learn that the
%surface density of the solid is about $\Sigma _d\propto a^{-1}$. For
%the location of Region II is between the Oort cloud and the
%scattered disk, we deduce that the number or the asteroids/comets in
%the quasi-static disk is about $2.5\times 10^9$. Under this
%assumption, the mass of the bodies in Region II is estimated to be
%0.0167 $M_{\oplus}$.

We make an analysis over other 28 groups of simulations and find
that the Damocloids mainly run back to two regions as shown in
Figure \ref{fig3}. We show our results in Table 2. A represents the
number of the test particles running back to Region II, while B
displays that in Region I. C and D illustrate the number of the test
particles that perturbed by the main planets and the sun
respectively. Thus $B=C+D$ is satisfied. The last column in Table 2
exhibits the ratio of the number in Region I and Region II. In order
to clearly illustrate the results, we define three new parameters:
$f_{\rm RegionI}$, $f_{\rm RegionII}$ means the probability of the
Damocloids from Region I and II, respectively, and \textit{f} is
expressed as
\begin {equation}
f=\frac{f_{\rm RegionI}}{f_{\rm RegionII}}.
\end {equation}

From the results in Table 2 and the definition of $f$, the
Damocloids can be further categorized into two types.

(1)Type I: $f<1$. Similar to the case of 2010 EJ104, over 50\% of
the test particles run back to Region II. It means that the
perturbation from Oort cloud is the main reason for the formation of
this type of Damocloids. 19 of 29 Damocloids could be grouped into
Type I, which are labeled in red in Figure \ref{fig1}.

(2)Type II: $f>1$. Different from type I, perturbation in the inner
solar system including the effect of the eight major planets and the
sun lead to the formation of Damocloids. While, the scattered disk
may be the most possible origin of the Damocloids. Ten Damocloids in
our samples are contained in this type, where are marked up in green
in Figure \ref{fig1}.

From the color of dots in Figure \ref{fig1}, we notice that Type I
and II have apparently marked off a boundary at about 15 AU. Based
on the statistic results over 29 Damocloids, the average possibility
of the Damocloids population originating from Region I and II are
about 65.5\% and 34.5\%, respectively.

%\begin{figure}
   %%\epsscale{1.2}
%   \plotone{f4.eps}
% \caption{Statistic results of all simulations. Panel (a), (b), (c) and (d)
%  shows the distribution of  possible origin for the Damocloids. Type I and II
%  mean the objects from two regions labeled in Figure 3. Type III indicates that the bodies
%  are mainly from Region I, while Type IV from Region II.}
% \label{fig4}
%\end{figure}

\begin{table*}[htp]
\begin{center}
\caption{Statistic results of the Damocloids. A represents the
number of the test particles that escaped from the inner solar
system; B displays the number of the test particles that perturbed
by the main planets or the sun; C means the number of the test
particles that perturbed by the main planets and D means the number
of the test particles that perturbed by the sun. Therefore, $B=C+D$.
The value of $f$ is got from equation 2. \label{tbl-2}}
\begin{tabular*}{8cm}{@{\extracolsep{\fill}}lccccc}
  \hline\noalign{\smallskip}
Name       & A   & B &C  & D &f \\
  \hline\noalign{\smallskip}
2010 NV1           & 923    &77  & 77   & 0 &0.083 \\
2010 GW147         & 897    &103 & 103  & 0 &0.115\\
2004 NN8           & 891    &109 & 109  & 0 &0.122\\
2010 JH124         & 809    &191 & 174  & 17&0.236 \\
2002 RP120         & 746    &254 & 245  & 9 &0.340\\
2000 AB229         & 731    &269 & 250  & 19&0.368 \\
2000 HE46          & 717    &283 & 277  & 6 &0.395  \\
2005 OE            & 713    &287 & 256  & 31&0.403  \\
1999 LD31          & 712    &288 & 280  & 8 &0.404\\
1997 MD10          & 695    &305 & 266  & 39&0.439 \\
2010 OM101         & 662    &338 & 320  & 18&0.511 \\
2010 EJ104         & 630    &370 & 346  & 24&0.587 \\
2009 YS6           & 625    &375 & 353  & 22&0.600  \\
2009 AU16          & 621    &379 & 344  & 35&0.610 \\
1998 WU24          & 601    &399 & 344  & 55&0.664  \\
2006 RJ2           & 588    &412 & 390  & 22&0.701  \\
1999 LE31          & 573    &427 & 419  & 8 &0.745\\
1999 XS35          & 571    &429 & 392  & 37 &0.751\\
2005 SB223         & 562    &438 & 389  & 49&0.779 \\
2000 DG8           & 427    &573 & 516  & 57&1.342  \\
2005 TJ50          & 412    &588 & 513  & 75 &1.427\\
2004 YH32          & 387    &613 & 490  & 123&1.584  \\
2005 NP82          & 353    &647 & 535  & 112&1.833  \\
2006 VW266         & 282    &781 & 557  & 161&2.546 \\
2009 FW23          & 243    &757 & 599  & 158&3.115 \\
2010 LG61          & 228    &772 & 581  & 191&3.386 \\
2010 OA101         & 154    &846 & 541  & 305 &5.494 \\
2007 VA85          & 126    &874 & 674  & 200&6.937 \\
2009 HC82          & 60     &940 & 807  & 133 &15.667\\
  \hline\noalign{\smallskip}
\end{tabular*}
\end{center}
\end{table*}

As mentioned in the first section, the HFCs are mainly from two
possible regions, the scattered disk or the inner Oort cloud.
According to our results, the Damocloids  may come from two regions
too, the scattered disk or the Oort cloud which will be perturbed
into the transient disk locating from 700 AU to 1000 AU. In this
sense, the same origin may imply the Damocloids may be the inactive
nuclei of the HFCs. In future with more carefully investigation will
get the detailed results.

\section{Conclusions and Discussions}
On the basis of the overall dynamical analysis, we underline that
two regions are likely to serve as birthplace for the Damocloids,
the scattered disk inside 100 AU (response to Region I) and Oort
cloud (respond to Region II). According to their possible origin,
they can be further classified into two types: Type I indicates that
the objects mainly come from the Oort cloud which respond to the
backward orbit to Region II; Type II suggests the population is
mostly from the scattered disk which will be perturbed by the major
planets or the sun .

In this work, we did not consider the effect of the outer solar
system. Thus, the evolution in the region with the distance from the
central star larger than 1000 AU is not clear. If the Damocloids
come from inner Oort cloud, we may briefly summarize the possible
routes for such objects: firstly, the bodies in Oort cloud might be
stirred by perturbation of the passing stars or the tidal effect by
the Galactic disk, then continuously moved inward to the
intermediate region (Region II), and finally ejected into the inner
solar system.

Additionally, the outer asteroid belt is another possible origin of
Damocloids. Bodies removed from this region are found to fall under
the gravitational influence of Jupiter that scatters them to large
heliocentric distance (\cite{Fer02}). It is then possible that some
scattered asteroids can return as Damocloids.

In this paper we only study the origin of the Damocloids from the
point of backward simulations. Thus, scattered disk and inner Oort
cloud are just two possible origins of Damocloids. But still there
are other possible origins such as the asteroid belt including the
main belt and the outer belt. It will be much elucidated for their
origin with more members of Damocloids observed in future.

\begin{acknowledgements}
We thank Zhaori Getu, Hong Renquan and Hu Longfei who have
greatly contributed to the asteroid survey observation. Z.H.B is
supported by the National Natural Science Foundation of China
(Grants No. 10503013, 10933004), J.J.H is grateful to the support by
the National Natural Science Foundation of China (Grants 10973044,
10833001), the Natural Science Foundation of Jiangsu Province, and
the Foundation of Minor Planets of Purple Mountain Observatory.
\end{acknowledgements}

\label{lastpage}


\begin{thebibliography}{99}
%% you can type \apj for ApJ, \aap for A&A, \apss for Ap&SS, etc. Please consult
%% the macro chjaa.cls. You can also find them in aasguide.tex (AASTeX for ApJ, AJ, PASP)
%% Please follow the format of ChJAA's reference list

\bibitem[Bailey \& Emel'yanenko 1996]{BE96} Bailey, M. E.,
\& Emel'yanenko, V. V., 1996, MNRAS, 278, 1087

\bibitem[Byl 1990]{Byl90}
Byl, J. 1990, \aj,  99, 1632

\bibitem[Chambers 1999]{Cham99}
Chambers, J. E. 1999, \mnras,  304, 793

\bibitem[Davies et al. 2001]{Davies01}
Davies, J. K., et al., 2001, Icarus, 150, 69

\bibitem[Dones et al. 2004]{Don}
Dones, L., Weissman, P., Levison, H. F., \& Duncan, M. 2004, in
Comets II, ed. M. C. Festou, H. U. Keller, \& H. A. Weaver (Tucson:
Univ. Arizona), 153

\bibitem[Duncan \& levison 1997]{DL97}
Duncan, M. J., \& Levison, H. F. 1997, Science, 276, 1670

\bibitem[Duncan 2008]{Duncan08}
Duncan, M. J. 2008, SSRv, 138, 109

\bibitem[Fernandez et al. 2002]{Fer02}Fernandez, J. A., Gallardo, T. \&
Brunini, A. 2002, Icarus, 159, 358.

\bibitem[Fouchard et al. 2006]{Fou06}
Fouchard, M.  et al. 2006,  CeMDA,   95, 299

\bibitem[Gladman 2005]{Glad05}
Gladman, B. 2005, Science,  307, 71

\bibitem[Harris et al. 2001]{Harris01}Harries, A. W. et al., 2001,
Icarus, 153, 332

\bibitem[Jewitt 2005]{Jewitt05}
Jewitt, D. 2005, AJ, 129, 530

\bibitem[Ji et al. 2005]{Ji05}
Ji, J.~H., Liu, L., Kinoshita, H., \& Li, G.Y.  2005, \apj, 631, 1191

\bibitem[Ji et al. 2011]{Ji11}
Ji, J.~H., Jin, S., \& Tinney, C.~G.\ 2011, \apjl, 727, L5


\bibitem[Levison et al. 2001]{Levison01}
Levison, H. F., Dones, L., \& Duncan, M. J., 2001,
\aj, 121, 2253

\bibitem[Levison et al. 2004]{LM04}
Levison, H. F., Morbidelli, A., \& Dones, L., 2004, \aj, 128, 2553

\bibitem[Levison et al. 2006]{Levison06}
Levison, H. F. 2006, Icarus, 184, 619

\bibitem[Levison et al. 2009]{Levison09}
Levison, H. F., Bottke, W. F., Gounelle, M., Morbidelli, A.,
Nesvorny, D., \& Tsiganis, K. 2009, \nat, 460, 364


\bibitem[Matese \& Whitmire 2011]{Matese}
Matese, J. J., \& Whitemire, D. P., 2011, Icarus, 211, 926

\bibitem[Morbidelli \& Levison 2004]{ML}
Morbidelli, A., \& Levison, H. F. 2004, \aj, 128, 2564


\bibitem[Morbidelli et al. 2008]{Mor08}
Morbidelli, A., Levison, H.F., \& Gomes, R. 2008, The Solar System
Beyond Neptune, 275


\bibitem[Moro-Martin 2008]{MM}
Moro-Martin, A. 2008, Proceedings of IAU Symposium No. 249, 347

\bibitem[Oort 1950]{oo}
Oort, J. H. 1950, Bull. Astron. Inst. Netherlands, 11, 91

\bibitem[Raymond et al. 2004]{Ray04}
Raymond S. N., Quinn T., \& Lunine J. I., 2004, Icarus, 168, 1

\bibitem[Toth 2006]{toth06}
Toth, I. 2006, Proceedings of IAU Symposium No. 229, 67


\bibitem[Weissman 1990]{Wei}
Weissman, P. R. 1990, \nat, 344, 825

\bibitem[Weissman \& Levison 1997]{WL97}
Weissman, P. R., \& Levison, H. 1997, \apjl,  488, L133

\bibitem[Zhao et al. 2008]{Zhao07}
Zhao, H. B., Yao, J. et al. 2008, Proceedings of IAU Symposium No.
248, 565

\bibitem[Zhao et al. 2009]{Zhao09}
Zhao, H. B., Lu, H., Zhaori, G. T, et al. 2009, Science in China
Series G: Physics, Mechanics \& Astronomy, 52, 1790

\bibitem[Zhao 2010]{Zhao10}
Zhao, H. B. 2010,  Acta Astronomica Sinica, 51, 324



\end{thebibliography}
\end{document}